\documentclass[prl,twocolumn,nofootinbib,preprintnumbers,amssymb,amsfonts,amsmath,superscriptaddress,showpacs,hyperref]{revtex4-1}

\usepackage{csquotes}
\usepackage{epsf,epsfig,amsmath,amssymb,dsfont}
\usepackage{amsthm,mathrsfs} 
\usepackage{graphicx} 
\usepackage{float}
\usepackage{slashed}
\usepackage{float}
\usepackage{hyperref} 
\usepackage[]{subfigure}
\usepackage{multirow}
\usepackage{mathtools}
\usepackage{color}
\usepackage[framemethod=TikZ]{mdframed}
\mdfdefinestyle{MyFrame}{%
    linecolor=blue,
    outerlinewidth=2pt,
    roundcorner=20pt,
    innertopmargin=\baselineskip,
    innerbottommargin=\baselineskip,
    innerrightmargin=20pt,
    innerleftmargin=20pt,
    backgroundcolor=gray!50!white}

\newcommand{\be}{\begin{equation}}
\newcommand{\ee}{\end{equation}}
\newcommand{\baln}{\begin{align}}
\newcommand{\ealn}{\end{align}}
\newcommand{\ben}{\begin{equation*}}
\newcommand{\een}{\end{equation*}}

\long\def\symbolfootnote[#1]#2{\begingroup%
\def\thefootnote{\fnsymbol{footnote}}\footnote[#1]{#2}\endgroup}

\usepackage{tikz}
\usetikzlibrary{calc}

\xdefinecolor{mycolor}{RGB}{62,96,111} 
\colorlet{bancolor}{mycolor}

\begin{document}

\title{Quantum Superposition of Massive Objects and the Quantization of Gravity}

\author{Alessio Belenchia} 
\email{alessio.belenchia@oeaw.ac.at}
\affiliation{Institute for Quantum Optics and Quantum Information (IQOQI), Boltzmanngasse 3 1090 Vienna, Austria.}

\author{Robert M. Wald}
\email{rmwa@uchicago.edu} 
\affiliation{Enrico Fermi Institute and Department of Physics, The University of Chicago, 5640 South  Ellis Avenue,  Chicago,  Illinois 60637,  USA}

\author{Flaminia Giacomini}
\email{flaminia.giacomini@univie.ac.at}
\affiliation{Institute for Quantum Optics and Quantum Information (IQOQI), Boltzmanngasse 3 1090 Vienna, Austria.}
\affiliation{Vienna Center for Quantum Science and Technology (VCQ), Faculty of Physics,\\
University of Vienna, Boltzmanngasse 5, A-1090 Vienna, Austria}

\author{Esteban Castro--Ruiz}
\email{esteban.castro.ruiz@univie.ac.at}
\affiliation{Institute for Quantum Optics and Quantum Information (IQOQI), Boltzmanngasse 3 1090 Vienna, Austria.}
\affiliation{Vienna Center for Quantum Science and Technology (VCQ), Faculty of Physics,\\
University of Vienna, Boltzmanngasse 5, A-1090 Vienna, Austria}

\author{\v{C}aslav Brukner}
\email{caslav.brukner@univie.ac.at}
\affiliation{Institute for Quantum Optics and Quantum Information (IQOQI), Boltzmanngasse 3 1090 Vienna, Austria.}
\affiliation{Vienna Center for Quantum Science and Technology (VCQ), Faculty of Physics,\\
University of Vienna, Boltzmanngasse 5, A-1090 Vienna, Austria}

\author{Markus Aspelmeyer}
\email{markus.aspelmeyer@univie.ac.at}
\affiliation{Vienna Center for Quantum Science and Technology (VCQ), Faculty of Physics,\\
University of Vienna, Boltzmanngasse 5, A-1090 Vienna, Austria}

\begin{abstract}
We analyse a gedankenexperiment previously considered by Mari et al.~\cite{Mari:2015qva} that involves quantum superpositions of charged and/or massive bodies (``particles'') under the control of the observers, Alice and Bob. In the electromagnetic case, we show that the quantization of electromagnetic radiation (which causes decoherence of Alice's particle) and vacuum fluctuations of the electromagnetic field (which limits Bob's ability to localize his particle to better than a charge-radius) both are essential for avoiding apparent paradoxes with causality and complementarity. We then analyze the gravitational version of this gedankenexperiment. We correct an error in the analysis of Mari et al.~\cite{Mari:2015qva} and of Baym and Ozawa~\cite{baym2009two}, who did not properly account for the conservation of center of mass of an isolated system. We show that the analysis of the gravitational case is in complete parallel with the electromagnetic case provided that gravitational radiation is quantized and that vacuum fluctuations limit the localization of a particle to no better than a Planck length. This provides support for the view that (linearized) gravity should have a quantum field description.

\end{abstract}
\maketitle

\section{I. Introduction}

An understanding of the fundamental nature of gravity and spacetime remains one of the most significant open issues in theoretical physics. The lack of a background spacetime structure in general relativity---the spacetime metric itself is the dynamical variable---makes it impossible to formulate a quantum theory of gravity by simply applying standard procedures that work for other fields. Although one can formulate an entirely satisfactory quantum field theory of linearized gravity---it is just a massless spin-2 field---severe difficulties arise when one attempts to go significantly beyond this description. Thus, there have been suggestions that gravity/spacetime could be fundamentally classical, or that its marriage with quantum mechanics requires a radical change of perspective on quantization~\cite{Hossenfelder:2010zj,Penrose2014}, or that quantization of gravity could be an ill-posed question in the first place~\cite{dyson2013graviton}---although there also have been many arguments given for the necessity of a quantum description of gravity~\cite{bronstein2012republication,PhysRevLett.47.979,eppley1977necessity,PhysRevD.73.064025,PhysRevD.73.064025,0264-9381-25-15-154010,giampaolo2018entanglement}.

In order to gain more insight into the quantum properties of gravity, it is helpful to consider the gravitational field associated with a quantum source, as already discussed by Feynman~\cite{cecile2011role,Zeh2011}. This is the basis of proposals for actual experiments employing macroscopic masses in superpositions~\cite{PhysRevLett.47.979,Anastopoulos:2015zta,Carlesso:2017vrw,Bahrami:2015wma,PhysRevA.71.024101,FORD1982238,kafri2013noise,Kafri:2014zsa,altamirano2016gravity}. The main aim of these works is to rule out semi-classical gravity as an exact theory~\cite{kiefer2007quantum,1367-2630-16-11-115007}, which would treat the gravitational field as classical even when the source is in a macroscopic superposition at different locations---in contrast with the expectations of standard quantum mechanics that a mass in superposition would generate a quantum superposition of gravitational fields. More recently, in~\cite{Bose:2017nin,Marletto:2017kzi} a novel way to witness entanglement due to solely the gravitational interaction was proposed. 
The authors use a gravitationally induced phase shift between two previously independent masses, both in superposition of different locations, which acts fully analogous to an entangling CSIGN gate~\cite{nielsen2002quantum}. They propose to witness the entanglement through correlation measurements between additional spin degrees of freedom. The claim is that, if entanglement between the spins of the two masses is certified then gravity should be a \textit{quantum coherent mediator} (see also~\cite{giampaolo2018entanglement,hall2018two}).

However, as stressed already in~\cite{Anastopoulos:2015zta,Anastopoulos:2018drh}, all the previous proposals\footnote{With the notable exception of~\cite{FORD1982238}, in which a dynamical version of the Page--Geilker scenario is considered.} can be accounted for by just considering the (non-local) gravitational potential in the Schr\"odinger equation describing the two particles, without any reference to dynamical degrees of freedom of the gravitational field. This has led the authors of~\cite{Anastopoulos:2018drh} to argue that, even if successful in witnessing entanglement, experiments like~\cite{Bose:2017nin,Marletto:2017kzi} would say nothing about the quantum nature of the gravitational field.    

In this work we provide a different conclusion by revisiting a gedankenexperiment previously considered by ~\cite{Mari:2015qva}, which is very similar in its essential aspects to one introduced earlier by~\cite{baym2009two}. We first analyze the electromagnetic version of this gedankenexperiment and emphasize that the quantum nature of the electromagnetic field is essential to maintain a fully consistent description. We then show that the analysis of the gravitational version of this gedankenexperiment follows in complete parallel to the electromagnetic case. In the course of our analysis of the gravitational version, we correct an important error appearing in~\cite{Mari:2015qva} and~\cite{baym2009two}, where the conservation of the center of mass of an isolated system was not properly taken into account. We find that the quantum nature of the gravitational field---both with regard to the quantization of gravitational radiation and the impossibility of localization to better than a Planck length---is essential for avoiding inconsistencies in the behavior of this system. This weakens the claim of~\cite{Anastopoulos:2018drh} that it is impossible to say anything about the necessity of quantized dynamical degrees of freedom of gravity with table top experiments like~\cite{Bose:2017nin,Marletto:2017kzi}.    

In Sec.II~\ref{II} we describe the gedankenexperiment of~\cite{Mari:2015qva}. We analyze the electromagnetic version of this experiment in Sec.III and the gravitational version in Sec.IV. Our results are summarized in section V~\ref{V}. 

We will work in units with $\hbar=c=1$.


\section{\label{II}II. The Gedankenexperiment of Mari et al.}
Consider two parties Alice and Bob, at a distance $D$ from each other, each controlling a charged and/or massive body, with charges $q_{A}$ and $q_{B}$ and masses $m_{A}$ and $m_{B}$, respectively. In the electromagnetic version of this gedankenexperiment, we will ignore all gravitational effects; in the gravitational version, we will put $q_A = q_B = 0$ and consider the gravitational effects. Since we will be interested only in the center of mass degrees of freedom of the bodies, we will hereafter refer to these bodies as ``particles,'' but we emphasize that these particles need not be elementary particles (or atoms or molecules) and thus they may have large charge and/or mass. We assume that Alice's particle also has spin and that, in the distant past, she sent her particle through a Stern-Gerlach apparatus, putting it in an equal superposition, $\frac{1}{\sqrt{2}} (|L \rangle_A|\downarrow\rangle_A + | R \rangle_A|\uparrow\rangle_A )$, of states $| L \rangle_A $ and $| R \rangle_A $ spatially separated by distance $d$. 
We assume that Alice did this separation process adiabatically, so negligible (electromagnetic or gravitational) radiation was emitted. On the other hand, Bob's particle is initially held in a trap with a sufficiently strong confining potential so that any influences of Alice's particle on the state of Bob's particle are negligible. 

At a pre-arranged time, $t=0$, Bob makes a choice of either releasing his particle from the trap or leaving it in the trap. If he releases his particle, then his particle will react to the electromagnetic or gravitational influence of Alice's particle, which will depend on the two amplitudes corresponding to the states $|L\rangle_A,\, |R\rangle_A$. Bob's particle will thereby become correlated with Alice's, with location of the center of mass of Bob's particle getting correlated with the location of Alice's particle. Let $T_B$ denote the time at which Bob completes his experiment, at which time the center of mass displacement of the different possible locations of his particle will be denoted $\delta x$. If $\delta x$ is sufficiently large, the location difference will make the possible states of Bob's particle nearly orthogonal, so his particle will be nearly maximally correlated with Alice's, and thus Alice's particle will be in a highly mixed state. In other words, Bob has acquired maximal "which-path" information about Alice's particle.

\begin{figure}[h!]
\centering
\includegraphics[width=0.35\textwidth]{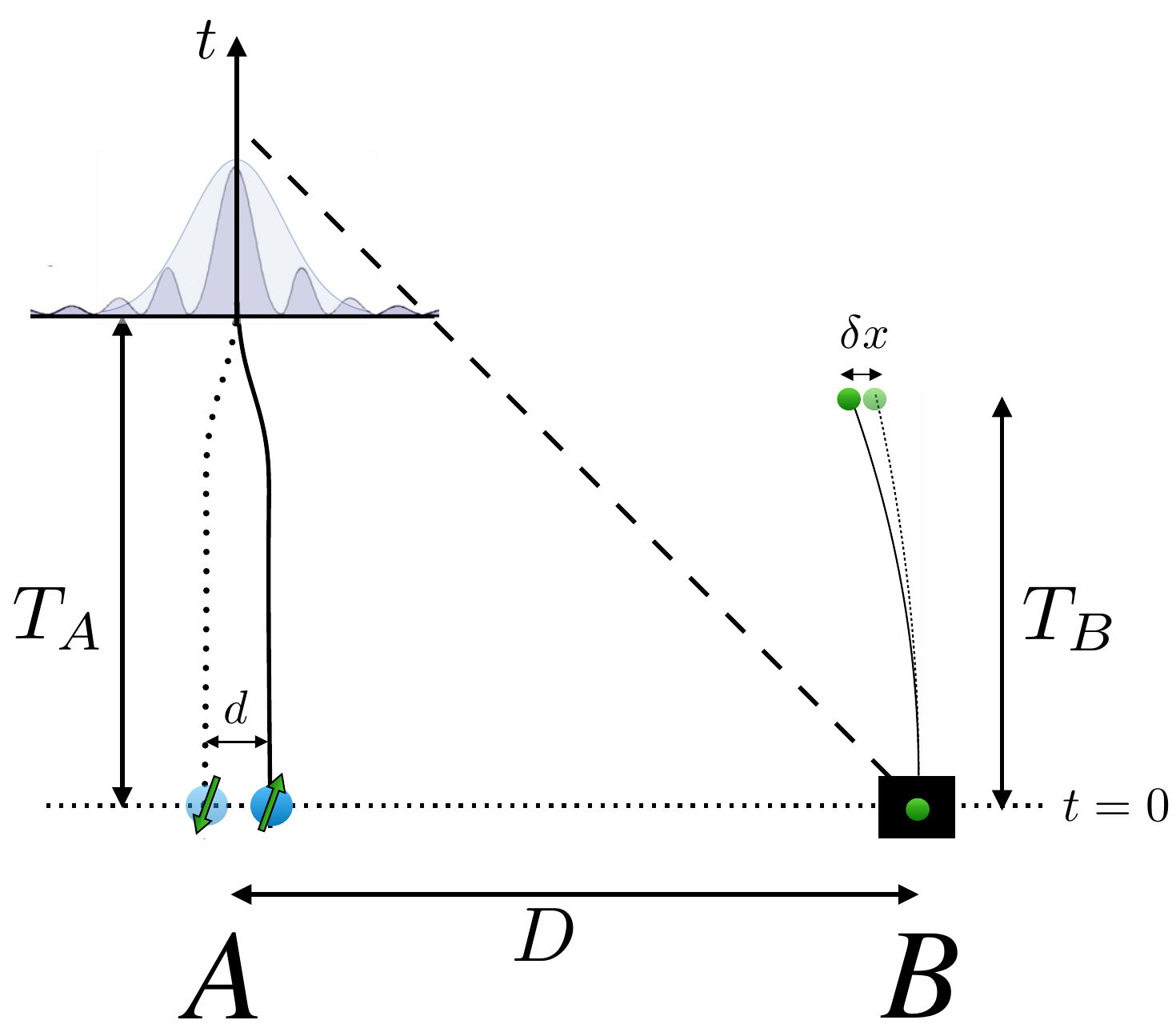}
 \caption{Arrangement of the gedankenexperiment. Alice's particle is prepared in a spatial superposition with separation $d$ while Bob's particle, at distance $D\gg d$, is initially localized by a trap. At the start of the protocol Bob can decide whether or not to release his particle from the trap, while Alice starts to recombine the paths of her particle. (When dividing and recombining the paths of her particle, Alice uses Stern-Gerlach devices, as discussed in~\cite{Mari:2015qva,Bose:2017nin}.) If Bob can acquire which-path information in a time $T_B<D$ and Alice recombines the superposition in a time $T_A<D$ without emitting radiation, then inconsistencies with causality or complementarity arise: Assuming complementarity holds, Alice could, by testing the coherence of her state (e.g., by measuring the additional spin degree of freedom in an appropriate basis, looking for the interference pattern, etc.) determine whether or not Bob opened the trap, in violation of causality. Alternatively, assuming causality holds, Alice could maintain the coherence of her state while which-path information has been acquired by Bob, in violation of complementarity.}
 \label{protocol}
\end{figure}

In the meantime, beginning also at $t= 0$, Alice sends her particle through a ``reversing'' Stern-Gerlach apparatus (similar to the experimental proposal in~\cite{Bose:2017nin}), in such a way that if her particle had remained unentangled (and thus in a pure state), she could successfully perform an interference experiment. She completes this process in time $T_A$. The arrangement of this gedankenexperiment is shown in Fig.~\ref{protocol}.

The most interesting case to consider is the one where $T_B < D$ and $T_A < D$ (as illustrated in Fig.~\ref{protocol}), so that Alice and Bob are spacelike separated from each other during the entire time when Bob's particle is (possibly) untrapped and Alice recombines her particle. This appears to lead to a paradoxical situation: If Bob opens the trap and can obtain ``which-path'' information from the behavior of his particle---which he should be able to do to at least to some degree---then, by the principle complementarity, Alice's particle must at least partially decohere, and she will then fail in her attempt to obtain a pure final state. On the other hand, if Bob left his particle in the trap, Bob's particle will remain in its (pure) ground state and cannot entangle with Alice's particle, so it might appear that Alice should be able to succeed in obtaining a pure final state. However, if this were the case, then the final state of Alice's particle would depend upon what Bob did, and Alice and Bob would be able to perform superluminal communication. We emphasize that if Bob can in any way influence the purity of Alice's final state when $T_B < D$ and $T_A < D$, then some degree of superluminal communication would be enabled. Conversely, if we assume that no superluminal communication is possible, i.e. that causality holds, then Bob's acquisition of ``which-path'' information without affecting the state of Alice's particle would appear to violate complementarity. This is analogous to the situation analyzed in~\cite{baym2009two}\footnote{See also Hardy's discussion of a similar gedankenexperiment, which he traces back to Y. Aharonov, in~\cite{schlosshauer2011elegance}.}. 
Thus it might appear that complementarity and causality cannot both hold.

We now analyze the electromagnetic version of this gedankenexperiment and show that a proper treatment of it leads to fully consistent results.
Although our discussion will be different in various respects and we will emphasize different aspects of some of the issues, our results in the electromagnetic case will be compatible with the previous analyses of \cite{Mari:2015qva} and \cite{baym2009two}.


\section{\label{III} III. Analysis of the Electromagnetic Version}

In our analysis of the electromagnetic gedankenexperiment, we will treat the particles of Alice and Bob via nonrelativistic quantum mechanics---which should be a good description if they are sufficiently massive and slow---and treat the electromagnetic field as a (relativistic) quantum field. 
After Alice sends her particle through the Stern-Gerlach apparatus, her particle will be entangled with its own electromagnetic field, but we assume that it is not entangled with anything else. Hence, the state of the system at time $t=0$---i.e., just before Bob opens the trap and just before Alice starts to recombine her particle---is described by
\begin{equation} \label{initial}
|\Psi \rangle =  \frac{1}{\sqrt{2}}\bigg[| L\rangle_A|\downarrow\rangle_A | \alpha_L \rangle_F+| R \rangle_A |\uparrow\rangle_A| \alpha_R \rangle_F \bigg] \otimes |\psi_0\rangle_B \, .
\end{equation}
Here $| L\rangle_A$ is the component of Alice's particle that ``went to the left'' under the Stern-Gerlach splitting, whereas $| R\rangle_A$ is the component of Alice's particle that ``went to the right''. The states $| \alpha_L \rangle_F$ and $| \alpha_R \rangle_F$ are the corresponding states of the electromagnetic field. More precisely, $| \alpha_L \rangle_F$ is assumed to be the coherent state of the electromagnetic field associated with the retarded solution with charge density $\rho_L = q_A |\psi_L|^2$ and current density $\vec{j}_L = q_A/m_A {\rm Im} (\bar{\psi}_L \vec{\nabla} \psi_L)$, where $\psi_L$ is the Schr\"odinger wavefunction of the state $| L\rangle_A$~\cite{unruh2000false,PhysRevA.68.062106,PhysRevA.63.032102,PhysRevD.47.5571,PhysRevA.56.1812}. Finally, $|\psi_0\rangle_B$ denotes the ground state of Bob's particle when it is in the trap.

We note that it will typically be the case that $|\langle \alpha_L | \alpha_R \rangle_F| \ll 1$, so in this sense, Alice's particle will have decohered at $t=0$, before Bob releases his particle and before she attempts to recombine her particle. However, as discussed by Unruh~\cite{unruh2000false}, this is a ``false decoherence.'' If Bob keeps his particle in the trap and Alice recombines her particle adiabatically, she will have no difficulty, in principle, in restoring the coherence, as the field ``follows'' the particle and recombines itself when the particle is recombined. 

We are interested in determining the effects of Bob opening the trap on the decoherence of Alice's particle. Our aim is to obtain a qualitative understanding of what phenomena play key roles as well as to obtain a semi-quantitative understanding of the magnitude of these phenomena. Thus, we will be content with making only rough, order of magnitude estimates and we will routinely discard numerical factors of order unity when making these estimates. As previously stated, we will work in units with 
$\hbar=c=1$.

First, we note that Alice's and Bob's particles do not interact with each other directly. Rather they each interact with the local electromagnetic field. Consequently, since the quantum electromagnetic field propagates in an entirely causal manner, it is clear that when $T_A < D$ and $T_B <D$, it is impossible\footnote{Here we neglect/ignore any effects of superluminal ``wavepacket spreading,'' which can occur in our nonrelativistic treatment of the particles via nonrelativistic quantum mechanics.} for anything that Bob does to influence the outcome of Alice's experiment. Thus, there can be no violation of causality. Second, since we analyze the gedankenexperiment using quantum theory, we can assign a quantum state to the system at all times. Since complementarity is a feature of all quantum states, there can be no violation of complementarity. Nevertheless, it is of interest to understand clearly how the paradoxes of Section II are being resolved, i.e. {\em why} Bob's actions do not influence the decoherence of Alice's particle when $T_A < D$ and {\em how} they can influence this decoherence when $T_A > D$. 

There are two properties of the quantum electromagnetic field that play a crucial role in our analysis. The first is the presence of vacuum fluctuations. (Similar fluctuations occur in all physically reasonable states.) When averaged over a spacetime region of (space and time) dimension $R$---recall that we have set $c=1$---the magnitude of the vacuum fluctuations of the electric field will be of order\footnote{Here, $\Delta E = \sqrt{\langle [E(f)]^2\rangle}$, where $E(f)$ is the electric field smeared with a smooth function $f$ with support in a region of size $R$ that is nearly constant in this region and normalized so that $\int f = 1$. Eq.~\eqref{flu} follows from the fact that the two-point correlation function of the vector potential behaves as $1/\sigma$ ---where $\sigma$ denotes the squared geodesic distance between the points---and the electric field is constructed from one derivative of the vector potential.}~\cite{bohrros1933}
\begin{equation}\label{flu}
\Delta E \sim 1/R^2 \, .
\end{equation}
When averaged over a worldline for a timescale $R$, the electric field will randomly fluctuate by this magnitude. The classical motion of a free, nonrelativistic particle of charge $q$ and mass $m$ will be influenced by this electric field according to Newton's second law, $m \ddot{x} = q E$. Integrating this equation, we find that the vacuum fluctuations of the electromagnetic field will displace the position of a classical free particle over the timescale $R$ by the amount
\begin{equation}
\Delta x \sim q/m 
\end{equation}
independently of $R$. Thus, as a consequence of vacuum fluctuations, a classical free particle cannot be localized to better than its {\em charge-radius}\footnote{Re-inserting $\hbar$ and $c$ the charge radius is given by $\hbar q/(m c q_P)$, where $q_P=\sqrt{\hbar c}$ is the Planck charge. In the text $q_P=1$.} $q/m$. We assume that the same must be true for a quantum free particle. Note that the charge-radius localization limit is more stringent than the localization limit given by the Compton wavelength, $1/m$, only when\footnote{For an electron, we have $q = \sqrt{\alpha} \sim 10^{-1}$, so we must use composite bodies to achieve $q > 1$.}  $q > 1$. However, it should be possible to evade the Compton wavelength localization limit by using relativistic bodies, whereas the charge radius localization limit is a fundamental limit arising from the quantum nature of the electromagnetic field.

The inability to localize a particle to better than its charge radius has the consequence of limiting Bob's ability to entangle his particle with Alice's. In order for significant entanglement to be achieved  after Bob releases his particle from the trap, it is necessary that the difference in the electromagnetic fields resulting from the different components of the wavefunction of Alice's particle be large enough to produce a displacement
\begin{equation} \label{chrad}
\delta x > q_B/m_B 
\end{equation}
in the motion of Bob's particle. 

The second key property of the quantum electromagnetic field is the existence of quantized electromagnetic radiation. When Alice recombines her particle with her ``reverse Stern-Gerlach'' apparatus, the effective dipole moment ${\mathcal D}_A = q_A d$ resulting from the difference in the retarded fields of the different components of her particle's wavefunction will be reduced to zero. If Alice is able to do the recombination sufficiently adiabatically, no radiation will be emitted, and $| \alpha_L \rangle_F$ and $| \alpha_R \rangle_F$ in \eqref{initial} will both adiabatically return to the vacuum state $| 0 \rangle_F$. If there are no influences from Bob's particle (or any environmental degrees of freedom), Alice then will be able to succeed in her coherence experiment; the final state of her particle will be pure. However, if Alice has to complete her experiment within time $T_A$, she may not be able to perform the recombination adiabatically and $| \alpha_L \rangle_F$ and $| \alpha_R \rangle_F$ will not return to the vacuum. If the resulting final states of the electromagnetic field differ by $\gtrsim 1$ photon, then they will be (nearly) orthogonal. This will cause her attempt at recoherence to fail.

Let us now estimate the size of the above two effects. First, when Bob opens his trap for time $T_B < D$, he experiences only the static electric fields associated with the different components of Alice's particle. His ability to obtain ``which-path'' information will rest entirely on his ability to detect the effective dipole moment ${\mathcal D}_A$ resulting from the difference in the retarded fields of the different components of Alice's particle's wavefunction. The electric field difference associated with this effective dipole moment is $E \sim {\mathcal D}_A/D^3$. If Bob's particle is released for a time $T_B$, the separation $\delta x$ between Bob's particle's center of mass positions will be given by
\begin{equation}\label{deltaxem}
    \delta x \sim \frac{q_B E}{m_B} T_{B}^2 = \frac{q_{B}}{m_{B}}\frac{{\mathcal D}_A}{D^3}T_{B}^2 \, .
\end{equation}
Comparing with \eqref{chrad}, we see that Bob will be able to obtain significant ``which-path'' information concerning Alice's particle if and only if
\begin{equation} \label{Binfo}
    \frac{{\mathcal D}_A}{D^3}T_{B}^2 > 1 \, .
\end{equation}

On the other hand, the amount of entangling radiation that is emitted by Alice's particle during the recombination can be estimated as follows. Each component of Alice's particle's superposition corresponds to a charge-current source $j^{\mu}_{L(R)}$, which is assumed to generate the corresponding coherent states $|\alpha_{R(L)}\rangle$. In general, the overlap can be written as  $\langle\alpha_{R}|\alpha_{L}\rangle=\langle 0|\alpha_{L}-\alpha_R\rangle$, where $|\alpha_{L}-\alpha_{R}\rangle$ is the coherent state generated by the difference $j_L-j_R$ (see~\cite{PhysRevA.68.062106,PhysRevA.63.032102}). The latter corresponds to the effective dipole ${\mathcal D}_A (t)$. Classically, the energy flux radiated by a time dependent dipole is $\sim (\ddot{\mathcal D}_A)^2$, so the total energy, $\mathcal E$, radiated when Alice ``closes her dipole'' will be given by
\begin{equation} 
\mathcal E \sim \left(\frac{{\mathcal D}_A}{T^2_A}\right)^2 T_A
\end{equation}
In quantum theory, this energy will appear in the form of photons of frequency $\sim 1/T_A$. Therefore, the number of photons in the state  $|\alpha_{L}-\alpha_{R}\rangle$ will be of order
\begin{equation} 
N \sim \left(\frac{{\mathcal D}_A}{T_A}\right)^2 
\end{equation}
Therefore, Alice can avoid emitting entangling radiation if and only if
\begin{equation}\label{dipoleem}       
{\mathcal D}_A < T_A
\end{equation}
This result has been previously obtained in~\cite{PhysRevD.47.5571,PhysRevA.56.1812,PhysRevA.68.062106,PhysRevA.63.032102} with varying degrees of detail and different techniques. 

We are now in a position to analyze the outcomes of the gedankenexperiment of section II in the various possible cases. First, consider the main case of interest, namely $T_A < D$ and $T_B < D$, so that Alice closes her superposition and Bob opens the trap in spacelike separated regions. This case divides into two subcases according to whether ${\mathcal D}_A < T_A$ or ${\mathcal D}_A > T_A$. If ${\mathcal D}_A < T_A$, then according to \eqref{dipoleem}, Alice can close her superposition without emitting entangling radiation. But, since ${\mathcal D}_A < T_A < D$, it follows from \eqref{Binfo} that Bob is unable to acquire ``which-path'' information in time $T_B < D$. Thus, Alice can successfully recohere her particle, and Bob can do nothing to stop her. On the other hand, if ${\mathcal D}_A > T_A$, then Alice's particle will necessarily emit entangling radiation, and her recoherence experiment will fail for this reason. Bob's particle can also obtain ``which-path'' information. The state of his particle will thereby be correlated with the state of Alice's particle. But Bob is entirely an ``innocent bystander'' in the decoherence of Alice's particle. When he opens the trap, his particle is merely entangling with the electromagnetic field that was already entangled with Alice's particle. He does not contribute in any way to Alice's particle's decoherence. 

By contrast, it is interesting to consider the case where we drop the limitation $T_A < D$. In particular, suppose that\footnote{Note that ${\mathcal D}_A > D$ requires $q_A \gg 1$, since we assume that $d \ll D$.} ${\mathcal D}_A > D$, so that, according to \eqref{Binfo}, Bob would acquire ``which-path'' information during the time $T_B < D$ if he releases his particle from the trap. However, suppose that Alice takes time $T_A > {\mathcal D}_A$ to close her superposition, so that she does not emit any entangling radiation. In that case, if Bob did not release his particle from the trap, Alice can successfully recohere her particle. However, if Bob did release his particle, it will get entangled with Alice's and her recoherence experiment will fail. There is no causality issue with this because we have $T_A > {\mathcal D}_A > D$. But it is interesting that what would have been a ``false decoherence'' of Alice's particle resulting from its entanglement with its own electromagnetic field becomes a true decoherence if Bob gets into the act. No matter how slowly she recombines her particle, Alice will be unable to undo her initial entanglement with electromagnetic states $| \alpha_L \rangle_F$ and $| \alpha_R \rangle_F$, which will be transferred to entanglement with Bob's particle. In this case, Bob is no longer an ``innocent bystander''; he is responsible for the failure of Alice's recoherence experiment. Interestingly, while Bob is carrying out his experiment, he has no idea whether he will be an innocent bystander or the culprit responsible for destroying the coherence of Alice's particle.

Finally, we consider what would happen if Alice tries to collect the radiation emitted by the particle and then combine the radiation with her particle in a recoherence experiment. The case of main interest is $D < {\mathcal D}_A$, since otherwise Bob would not be able to gain ``which-path'' information in the allotted time in any case. In order for Alice to collect the radiation, she will need the equivalent of a spherical mirror surrounding her apparatus. She can either (I) have this mirror be present during the entire experiment or (II) erect this mirror over a time $T_M < D$ beginning near $t=0$. She also has the choice of placing the mirror at distance (a) $R_M < D$ or (b) $R_M > D$ from her. In case (Ia), Alice will be able to successfully perform the recoherence experiment, but the mirror will shield the effective dipole from Bob, who will not be able to gain any ``which-path'' information. In cases (Ib) and (IIb), Bob will be able to gain ``which-path'' information and will be responsible for the decoherence of Alice's particle, just as in the case $T_A > D$ discussed in the previous paragraph. Finally, in the case (IIa), the erection of the mirror in time $T_M < D < {\mathcal D}_A$ will produce a time changing dipole moment, which will result in the emission of entangling photons to infinity just as in the case $T_A < D < {\mathcal D}_A$ discussed above. Again, no difficulties with causality or complementarity arise.

In summary, our analysis of this gedankenexperiment yields entirely consistent results that are compatible with causality and complementarity\footnote{Our argument also encompasses the set-up described in~\cite{baym2009two} which slightly differs from Fig.\ref{protocol}.}. We emphasize that both vacuum fluctuations of the electromagnetic field and the quantization of electromagnetic radiation were essential for obtaining this consistency. Without vacuum fluctuations, in the case ${\mathcal D}_A <  D$, Bob should be able to obtain ``which-path'' information in time $T_B < D$, resulting in a violation of causality if he influences Alice's state and a violation of complementarity if he doesn't. Similarly, without quantized radiation, in the case where ${\mathcal D}_A > D$, Alice would be able to recohere her particle in time $T_A<D$ (if not influenced by Bob), but Bob can obtain significant ``which-path'' information in time $T_B < D$. Again, this would lead to a violation of causality or complementarity.


\section{\label{IV} IV. Analysis of the Gravitational Version}

We now consider the gravitational version of the gedankenexperiment of Sec. II. We set $q_A = q_B = 0$ but now take into account the gravitational interaction. We continue to treat the particles via nonrelativistic quantum mechanics and we treat the (linearized) gravitational field as a quantum field. Our main aim is to show that, as in the electromagnetic version, the vacuum fluctuations of the gravitational field and the quantization of gravitational radiation are essential for the consistency of the analysis. 

The quantization of (linearized) gravitational radiation does not require any further discussion from us. However, we shall now briefly discuss the implications of vacuum fluctuations for the ``localization'' of a particle.
On account of the absence of background structure in general relativity, the ``location'' of a particle is not a well defined concept. The best one can do is consider the relative location of two bodies. Consider two bodies separated by a distance $R$. 
When averaged over a spacetime region of (space and time) dimension $R$, the magnitude of the vacuum fluctuations of the Riemann curvature tensor $\mathcal R$ should be of order\footnote{This follows from the fact that the correlation function of the linearized metric diverges as $l^2_P/\sigma$, where $\sigma$ denotes the squared geodesic distance between the points. Since the linearized Riemann tensor is constructed from two derivatives of the metric, the correlation function of the linearized Riemann tensor diverges as $l^2_P/\sigma^3$, yielding the estimate~\eqref{riem}.}
\begin{equation}\label{riem}
\Delta \mathcal R \sim l_P/R^3 
\end{equation}
where $l_P = \sqrt{G}$ is the Planck length. (Re-inserting $\hbar$ and $c$, we have $l_P = (G \hbar/c^3)^{1/2} \sim 10^{-35} {\rm m}$.) Integration of the geodesic deviation equation 
over time $R$ then yields the result that the two bodies should fluctuate in their relative position by an amount
\begin{equation}
\Delta x \sim l_P 
\end{equation}
independently of $R$ (and independently of the mass or other properties of the body). This leads to the conclusion that localization of any body cannot be achieved to better than a Planck length---a conclusion that has been previously reached by many authors; see~\cite{PhysRevLett.93.211101},~\cite{Garay:1994en,Hossenfelder:2012jw} and references therein. Thus, in our gedankenexperiment, in order for significant entanglement to be achieved after Bob releases his particle from the trap, it is necessary that the difference in the gravitational fields resulting from the different components of the wavefunction of Alice's particle be large enough to produce a displacement
\begin{equation} 
\delta x > l_P
\end{equation}
in the motion of Bob's particle. In the following, we will work in Planck units by setting $G=1$ (in addition to $\hbar = c = 1$), so $l_P = 1$.

We now estimate more quantitatively Bob's ability to obtain ``which-path'' information as well as the entanglement of Alice's particle with gravitons. Here, there is one very significant difference with the electromagnetic case: One might expect that, as in the electromagnetic case, the separation of Alice's particle into a superposition of different paths would produce an effective mass dipole moment, which would provide the leading order effect that Bob could use to entangle his particle with Alice's. However, this is not the case. Although Alice's particle, considered by itself, would lead to an effective mass dipole, the fact that she used a laboratory to do the separation cannot be neglected. Alice's total system consists of her particle and her laboratory (plus whatever her laboratory is attached to, such as the Earth). The stress-energy tensor of her total system is conserved. But conservation of stress-energy in a (nearly) flat spacetime implies that the center of mass of the total system moves on an inertial trajectory that cannot be altered no matter what internal changes take place in the system. In our case, this means that if Alice's particle ``moves to the left'' under the Stern-Gerlach splitting, then her laboratory must ``move to the right'' by just the right amount to keep the center of mass unchanged. Thus, Alice's particle must become entangled with her laboratory, in such a way that the state of her total system (ignoring the spin factors) is of the form
\begin{equation} 
|L \rangle | \beta_L \rangle  | \alpha_L \rangle + |R \rangle | \beta_R \rangle  | \alpha_R \rangle 
\end{equation}
where $| \beta_L \rangle$ and $| \beta_R \rangle$ are the corresponding laboratory states and $| \alpha_L(R) \rangle$ the gravitational field states (similarly to \eqref{initial}). This entanglement with the laboratory need not produce a significant decoherence---and this decoherence should be a ``false decoherence''~\cite{unruh2000false} in any case. The importance of including Alice's laboratory in the analysis is that the states $|L \rangle | \beta_L \rangle$ and $|R \rangle | \beta_R \rangle$ have exactly the same center of mass. Thus, the effective mass dipole resulting from the separation of Alice's particle vanishes. This point was overlooked in the analyses of ~\cite{baym2009two,Mari:2015qva}.

The vanishing of the effective mass dipole means that the dominant gravitational effect that Bob can use to obtain ``which-path'' information is the effective mass quadrupole\footnote{Assuming that the mass of Alice's laboratory is much greater than the mass of her particle, the contributions of her laboratory to the effective mass quadrupole may be safely neglected.}, ${\mathcal Q}_A$. The separation of Bob's components during time $T_B$ will now be given by
\begin{equation}\label{deltaxem}
    \delta x \sim  \frac{{\mathcal Q}_A}{D^4}T_{B}^2 \, .
\end{equation}
Thus, Bob will be able to obtain ``which-path'' information only when (recalling that we have set $l_P = 1$)
\begin{equation} \label{Binfo2}
 \frac{{\mathcal Q}_A}{D^4}T_{B}^2 > 1  \, .
\end{equation}

As in the electromagnetic case, Alice's particle will radiate when she performs the recombination, which will lead to some degree of decoherence. However, now the dominant entangling radiation will be quadrupole rather than dipole in nature. (The absence of dipole gravitational radiation is, of course, directly related to the conservation of center of mass.) The energy radiated while she ``closes her quadrupole'' will be given by
\begin{equation} 
\mathcal E \sim \left(\frac{{\mathcal Q}_A}{T^3_A}\right)^2 T_A \, .
\end{equation}
The corresponding number of gravitons is 
\begin{equation} 
N \sim \left(\frac{{\mathcal Q}_A}{T^2_A}\right)^2 
\end{equation}
Thus, Alice can avoid emitting entangling radiation only when\footnote{This discussion can be generalized to higher multipoles, as would be needed if, e.g., the effective quadrupole happened to vanish for a particular configuration. In this case equation~\eqref{deltaxem} is replaced by $ \delta x\sim (\mathcal{Q}^{(n)}_A/D^{n+2}) T_B^2$, where $\mathcal{Q}^{(n)}$ is the mass $2^n$th-multipole ($n\geq 2$). Similarly, eq.~\eqref{qurad} will be replaced by $\mathcal{Q}_{A}^{(n)}<T^{n}_A$. } 
\begin{equation} \label{qurad}
{\mathcal Q}_A < T^2_A \, .
\end{equation}

We now are in a position to analyze the various outcomes of the gravitational gedankenexperiment. As before, the main case of interest is $T_A < D$ and $T_B < D$. This case divides into two subcases, according to whether ${\mathcal Q}_A < T^2_A$ or ${\mathcal Q}_A > T^2_A$. In the first case, according to \eqref{qurad}, Alice can avoid emitting significant entangling radiation, but according to \eqref{Binfo2}, Bob will be unable to obtain significant ``which-path'' information in the allotted time. Thus, we find that Alice can recohere her particle within the allotted time, and this leads to no inconsistency with Bob's experiment. On the other hand, if ${\mathcal Q}_A > T^2_A$, Alice will necessarily emit entangling gravitational radiation, and her particle will decohere no matter what Bob does. Bob can obtain ``which-path'' information in time $T_B < D$ if he opens his trap---and the state of his particle will become correlated with Alice's---but he is merely transferring the entanglement with Alice's particle that was already present in the gravitational field to his particle. He is entirely an ``innocent bystander'' with regards to the reason why Alice's particle decohered.

The analysis of the other cases also follows in exact parallel with the electromagnetic version. We conclude that by treating the (linearized) gravitational field as a quantum field, we obtain an entirely consistent analysis of the gravitational version of this gedankenexperiment, which is fully compatible with causality and complementarity.


\section{\label{V} V. Summary and Conclusions}

We have carefully re-analyzed the gedankenexperiment of Mari et al.~\cite{Mari:2015qva}  in both its electromagnetic and gravitational versions.

In the electromagnetic case, we found that consistent results compatible with causality and complementarity are obtained, but that it was essential to take into account the quantum nature of the electromagnetic field both with regard to vacuum fluctuations (which limit Bob's ability to obtain ``which-path'' information) and the quantum properties of radiation (which cause Alice's particle to become entangled with photons if she performs the recombination too quickly). We then analyzed the gravitational case and found that exactly analogous results hold, with the substitution ``dipole'' $\to$ ``quadrupole.'' Again, the quantum nature of the (linearized) gravitational field played a crucial role in our analysis, both with regard to vacuum fluctuations and the quantum properties of radiation.

The main significance of our results is that it can be seen that the quantum properties of the gravitational field are essential for obtaining a consistent description of a system that otherwise should be well described by nonrelativistic quantum mechanics. We conclude that if Alice's and Bob's particles are well described by nonrelativistic quantum mechanics, then (linearized) gravity must possess the properties of a quantum field with regard to vacuum fluctuations and the quantum properties of radiation. Conversely, if one wishes to deny either of these quantum field properties of gravity, one must be prepared to make drastic modifications to the nonrelativistic quantum mechanics of massive particles (see e.g.~\cite{0264-9381-34-19-193002} and references therein).

We hope that our result helps to settle the current controversy whether the quantum nature of gravity is required in understanding the occurrence of gravitationally induced entanglement. However, we do not believe that our analysis offers any insights into the nature of a complete quantum theory of nonlinear gravity, except that its properties should be like those of an ordinary quantum field theory in the linearized limit.


\section*{Acknowledgement}
The authors would like to thank Vittorio Giovannetti, Andrea Mari, Giacomo De Palma, Miles Blencowe and Markus M\"uller for stimulating discussions and the reading of an early draft of the work. AB thanks Philipp H\"ohn, Luis C. Barbado, Eduardo Martin--Martinez, Matteo Carlesso, Giulio Gasbarri, Andr\'{e} Grossardt, Sougato Bose, Claus Kiefer, Matteo Lostaglio, Carlo Rovelli and Mauro Paternostro for stimulating discussions. CB thanks Borivoje Dakic and Tomasz Paterek, and MA Myungshik Kim and Gerard Milburn, for interesting discussions. AB wishes to acknowledge the hospitality of Queen's University Belfast where part of this work was carried out and the support of  STSM Grant from COST Action CA15220. AB, CB, FG, EC and MA acknowledge the support of the Austrian Academy of Sciences through  Innovationsfonds  Forschung,  Wissenschaft  und  Gesellschaft,  and the University of Vienna through the research platform TURIS. FG and EC acknowledge support from the the doctoral program ``Complex Quantum Systems" (CoQuS). This project has received funding from the European Union’s Horizon 2020 research and innovation programme under grant agreement No 766900 and from the European Research Council (ERC) under the European Union’s Horizon 2020 research and innovation programme (grant agreement No 649008). This publication was made possible through the support of a grant from the John Templeton Foundation. The opinions expressed in this publication are those of the authors and do not necessarily reflect the views of the John Templeton Foundation. The research of RMW was supported in part by NSF grants PHY 15-05124 and PHY18-04216 to the University of Chicago.


\bibliographystyle{JHEP}
\bibliography{references2.bib}


\end{document}